\documentclass[a4paper,USenglish]{lipics}
\usepackage{microtype}
\usepackage{here}       
\usepackage{mathrsfs}
\usepackage{yfonts}
\usepackage{multirow}
\usepackage[all]{xy}
\usepackage{latexsym}
\usepackage{color}
\usepackage{calc}
\usepackage{pgf}
\usepackage{multicol}  
\usepackage{framed}
\usepackage{fancybox}

\newcommand{\cardinality}[1]{\textit{card} \left( #1 \right)}
\newcommand {\calmind}{\hbox{\tiny\hbox{$\cal M$}}}
\newcommand {\EV}[2]{\ifmmode\hbox{$\mathbb Ev$}\left(#1,#2\right)\else$\hbox{$\mathbb Ev$}\left(#1,#2\right)$\fi}

\title{Easy  Proofs of  L\"owenheim-Skolem Theorems by Means of Evaluation Games} 
\titlerunning{AEasy  Proofs of  L\"owenheim-Skolem}

\author[1,2]{Jacques Duparc}
\affil[1]{Department of Information, Systems Faculty of Business and Economics\\
University of Lausanne, CH-1015 Lausanne Switzerland\\
  \texttt{Jacques.Duparc@unil.ch}}
\affil[2]{Mathematics Section,  School of Basic Sciences\\
Ecole polytechnique f\'{e}d\'{e}rale de Lausanne, CH-1015 Lausanne Switzerland \\
  \texttt{Jacques.Duparc@epfl.ch}}
\authorrunning{J. Duparc}

\Copyright{Jacques Duparc}

\subjclass{F.4.1 Mathematical Logic}

\keywords{Model theory, L\"owenheim-Skolem, Game Theory, Evaluation Game, First-Order Logic} 

\serieslogo{logo_ttl}
\volumeinfo
  {M. Antonia {Huertas}, Jo\~ao {Marcos}, Mar\'ia {Manzano}, Sophie {Pinchinat}, \\
  Fran\c{c}ois {Schwarzentruber}}
  {5}
  {4th International Conference on Tools for Teaching Logic}
  {1}
  {1}
  {27}
\EventShortName{TTL2015}

\begin{document}

\maketitle

\parindent=0em

\begin{abstract}We propose a proof of the downward L\"owenheim-Skolem that relies on  strategies deriving from evaluation games
 instead of the Skolem normal forms.
This proof is simpler, and  easily understood by the students, although  it requires, when defining the semantics of first-order logic 
 to introduce first a few notions inherited from game theory such as the one of an evaluation game. 
\end{abstract}

\section{Introduction}

Each mathematical logic course focuses on first-order logic. Once the basic definitions about syntax and semantics 
have been introduced and the notion of  the cardinality of a model has been exposed, sooner or later at least 
a couple of hours are dedicated 
to the L\"owenheim-Skolem theorem. This statement holds actually two different  results: the downward  L\"owenheim-Skolem theorem 
(\textbf{LS$\downarrow$}) and the upward L\"owenheim-Skolem theorem (\textbf{LS$\uparrow$}).

\begin {theorem} [Downward L\"owenheim-Skolem]Let $\mathcal{L}$ \label{DownwardLS}
	be a first-order  language,   $T$ some $\mathcal{L}$-theory, and $\kappa= \max \{\cardinality{\mathcal{L}}, \aleph_0\}$.

If  $T$ has a model of cardinality $\lambda > \kappa$, then  $T$ has a model of  cardinality~$\kappa$.
\end{theorem}

\begin {theorem} [Upward L\"owenheim-Skolem]
Let $\mathcal{L}$ be some first-order  language with equality,
	 $T$ some $\mathcal{L}$-theory, and $\kappa= \max \{\cardinality{\mathcal{L}}, \aleph_0\}$.

If  $T$ has an infinite model, then   $T$ has a model of  cardinality~$\lambda$, for any $\lambda > \kappa$.
\end{theorem}

The proof of the second theorem (\textbf{LS$\uparrow$}) is a simple exercise that relies on an easy  
application of the compactness theorem joined with a straightforward 
utilization of \textbf{LS$\downarrow$}. The proof of the first theorem is more involved and not that easy 
to understand for undergraduate students at EPFL. For some basic background and  notations we refer the reader to  \cite{cori2000mathematical1,marker2002model}.

The usual approach to proving  \textbf{LS$\downarrow$} goes through several steps which involve  reducing the original theory $T$ to 
another theory $T'$ on an extended language $\mathcal{L}'$ where all statements are in  Skolem normal form. Then obtaining some
  $\mathcal{L}'$-structure
 of the right cardinality that satisfies $T'$, from which one goes back to a model that satisfies $T$.
 
The burden of going through the Skolem normal forms is regarded as bothersome by the student.

We would like to advocate that the use of \emph{evaluation games} greatly simplifies the proof of \textbf{LS$\downarrow$}.  Of course, this 
requires  to talk about finite two-player perfect games and the related notions of player, strategies, winning strategies, etc. But 
it is worth the candle,  especially if these games are introduced for explaining the semantics of first-order logic. Anyhow, 
determining whether a first-order formula holds true  in a given structure is very similar to solving  the underlying evaluation game as   
put forward by Jaako Hintikka \cite{hintikka1979game}. We refer the readers unfamiliar with these notions 
 to \cite{vaananen2011models,mints2003games},  where the tight relations between logic and games  are disclosed.

\section{Evaluation games for first-order logic}

We noticed that when presented with both the classical semantics of first-order logic and the semantics that makes use of 
evaluation games, the audience gets much more involved in the second approach. 
Indeed, most students are more eager to solving games than to checking  whether a formula holds true.

Moreover, introducing first-order formulas not as (linear) sequences of symbols, but rather as trees (usually denoted as  
decomposition tree) makes it even easier to give evidence in support of the game-theoretical way of dealing with satisfaction.
The reason is twofold: 
\begin{enumerate}
  \item the whole arena of the evaluation is very similar to the tree decomposition of the formula. It has the very 
  same height and the same branching except when quantifiers are involved, where it depends on the cardinality of the domain of the model.
  \item the task of pointing out the occurrences of variables that are bound by a given quantifier -- in order to replace them by an element of 
  the domain chosen by one of the players -- is easily taken care of by taking the path  along  the unique branch that leads 
  from a given leaf of the tree where the   occurrence of the variable is situated, to the root of the tree. The first -- if any -- quantifier acting
   on this variable that is  encountered is the one that bounds it.

\end{enumerate}
We recall the definition of the evaluation game for first-order logic.

\begin{definition} Let $\mathcal{L}$ be a first-order language, ${\phi}$ some {\em closed} formula  
whose logical connectors are among $\{\neg,\vee,\wedge\}$, and  $\mathcal{M}$
 some $\mathcal{L}$-structure.

The evaluation game $\EV{\mathcal{M}}{{\phi}}$ is defined  as follows:\begin{enumerate}
\item there are two players, called {\em \textbf{V}erifier} and {\em \textbf{F}alsifier}.
{\em \textbf{V}erifier} ({\bf V}) has incentive to show that the formula holds in the $\mathcal{L}$-structure
 ($\mathcal{M}\models{\phi}$), whereas the goal of {\em \textbf{F}alsifier} ({\bf F})
 is to show that it does not hold ($\mathcal{M}\not\models{\phi}$).

The  moves of the players essentially consist of pushing a token down the tree decomposition of the formula ${\phi}$ -- as a way to choose 
 sub-formulas -- and must comply with the rules below:
\begin{center}
\scalebox{.8}{\fbox{\begin{tabular}[t]{|c|c|c|}
\hline
\ &&\\
 if the current position is\ldots   & whose turn \ldots & the game continues with\ldots\\\ &&\\\hline
\ &&\\
${\varphi_0 ~\vee~\varphi_1}$ & {\bf V} chooses $j\in\{0,1\}$ & ${\varphi_j}$\\
\ &&\\\hline
\ &&\\
${\varphi_0 ~\land~\varphi_1}$ & {\bf F} chooses $j\in\{0,1\}$ & ${\varphi_j}$\\
\ &&\\\hline
\ &&\\
${\neg\varphi}$ & {\bf F} and {\bf V} switch roles & ${\varphi}$\\
\ &&\\\hline
\ &&\\
${\exists x_i\varphi}$ & {\bf V} chooses $a_i\in|\mathcal{M}|$  & ${\varphi[a_i/x_i]}$\\
\ &&\\\hline
\ &&\\
${\forall x_i\varphi}$ & {\bf F} chooses  $a_i\in|\mathcal{M}|$  & ${\varphi[a_i/x_i]}$\\
\ &&\\\hline
\ &&\\
\ &&{\bf V} wins \\
${R(t_1,\cdots,t_k)_{[a_1/x_1,\cdots,a_n/x_n]}}$  & the game stops &
$\Longleftrightarrow$\\
\ &&$\mathcal{M}, a_1/x_1,\cdots,a_n/x_n\models{R(t_1,\cdots,t_k)}$\\
&&\\\hline
     \end{tabular}}}
	
\end{center}
\vspace{2ex}

\item The winning condition arises when the remaining formula becomes atomic, i.e. of the form
 ${R(t_1,\cdots,t_k)_{[a_1/x_1,\cdots,a_n/x_n]}}$. Notice that the rules guarantee that, since the initial formula is closed, 
 one always ends up with an atomic formula that does not contain any more variable as each of them has been replaced by some element 
 of the domain of the model
 \footnote{Formally we should not say that we replace each variable 
 $x_i$ by some element  $a_i$, but rather that we replace $x_i$  by some  {\em brand new}
constant symbol $\texttt{c}_{a_i}$, whose interpretation is precisely this element  $a_i$ and the formula we reach at the end is 
of the form ${R(t_1,\ldots,t_k)_{[\texttt{c}_{a_1}/x_1,\ldots,\texttt{c}_{a_n}/x_n]}}$.
}.

Player {\bf V} wins if ${R(t_1,\ldots,t_k)}$ is satisfied in the extended 
 $\mathcal{L}$-structure  $\mathcal{M}, a_1/x_1,\ldots,a_n/x_n$ ($\mathcal{M}, a_1/x_1,\ldots,a_n/x_n\models{R(t_1,\ldots,t_k)}$); {\bf F} wins otherwise.
\end{enumerate}
\end{definition}

\begin{example}
Suppose we have  
 a language that  contains a unary relation symbol $P$,  
a binary relation symbol $R$, and a unary function symbol \textit{f}. We then consider   
the formula  ${\forall x\Big(P(x)\vee\exists y R\big(f(x),y\big)\Big)}$ whose tree decomposition is
\begin{figure}[H]
\hfill\  
 \shadowbox{\includegraphics[width=2.6cm]{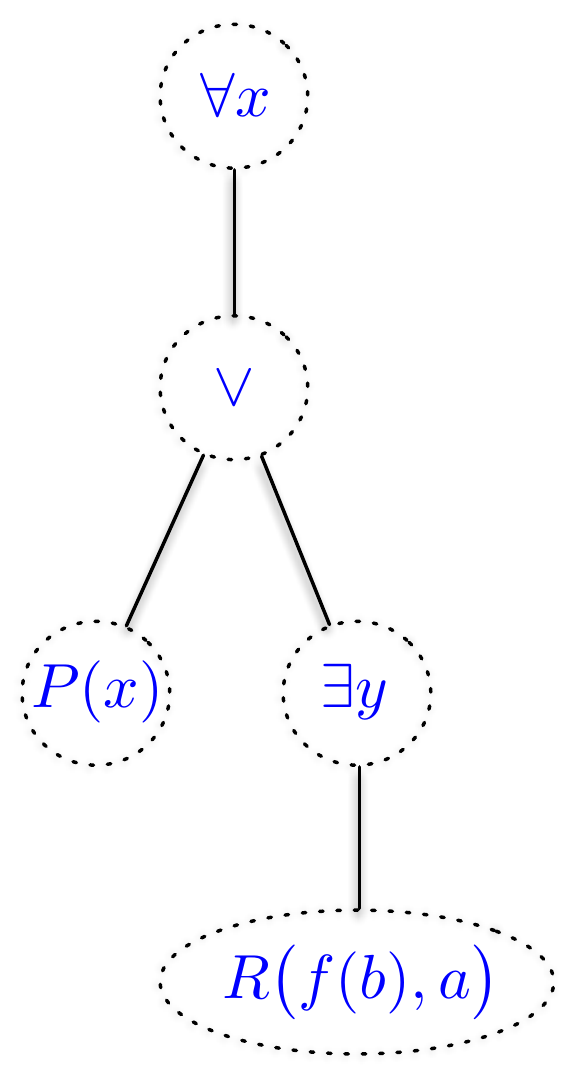}}
\hfill\  \end{figure}

 the model 
  $\mathcal{M}$ is defined by:\vspace{1ex}
     \begin{multicols}{3}
\begin{itemize}
\item $|\mathcal{M}|=\{a,b\}$,\vspace{.5ex}
\item $f^{\calmind}(a)=b$.

\item $R^{\calmind}=\{(b,a)\}$.\vspace{.5ex}
\item $f^{\calmind}(b)=a$ 

\item  $P^{\calmind}=\{b\}$,\vspace{.5ex}

   \end{itemize}

   \end{multicols}

The game tree that represents the arena for the evaluation game
 $\EV{\mathcal{M}}{{\forall x\left(P(x)\vee\exists y R(f(x),y)\right)}}$ is played on the arena represented by 
 the following game tree:
\begin{figure}[H]
\hfill\  
 \shadowbox{\includegraphics[width=8cm]{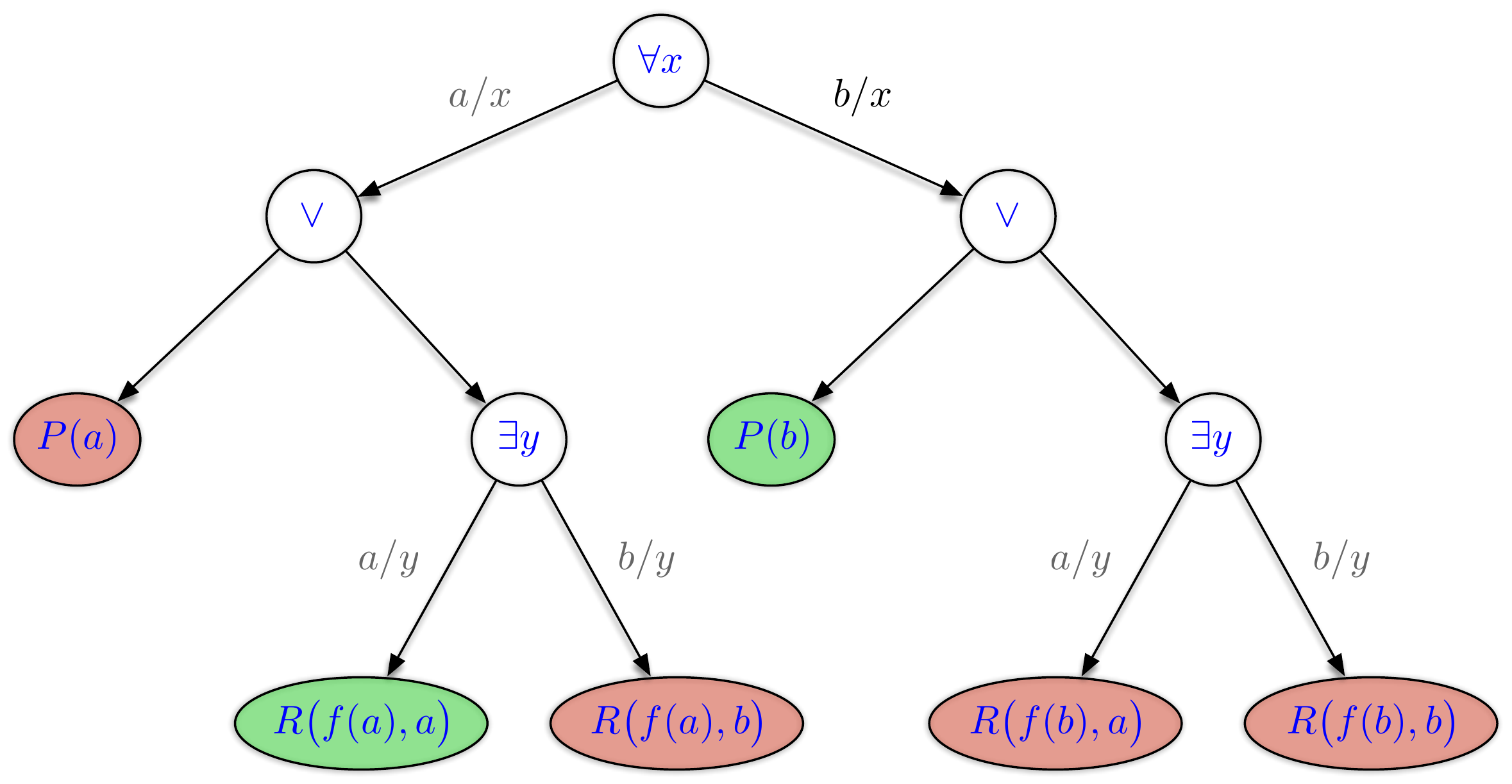}}
\hfill\  \end{figure}

The green leaves  are the ones where the atomic formula holds true in the model, and  the opposite for the red ones.

We then proceed by \emph{backward induction} and  
assign either the colour green or the colour red to every node depending on whether the 
{\em \textbf{V}erifier} or the {\em \textbf{F}falsifier} has a winning strategy if the game were to start from that particular node.
\begin{figure}[H]
\hfill\  
 \shadowbox{\includegraphics[width=8cm]{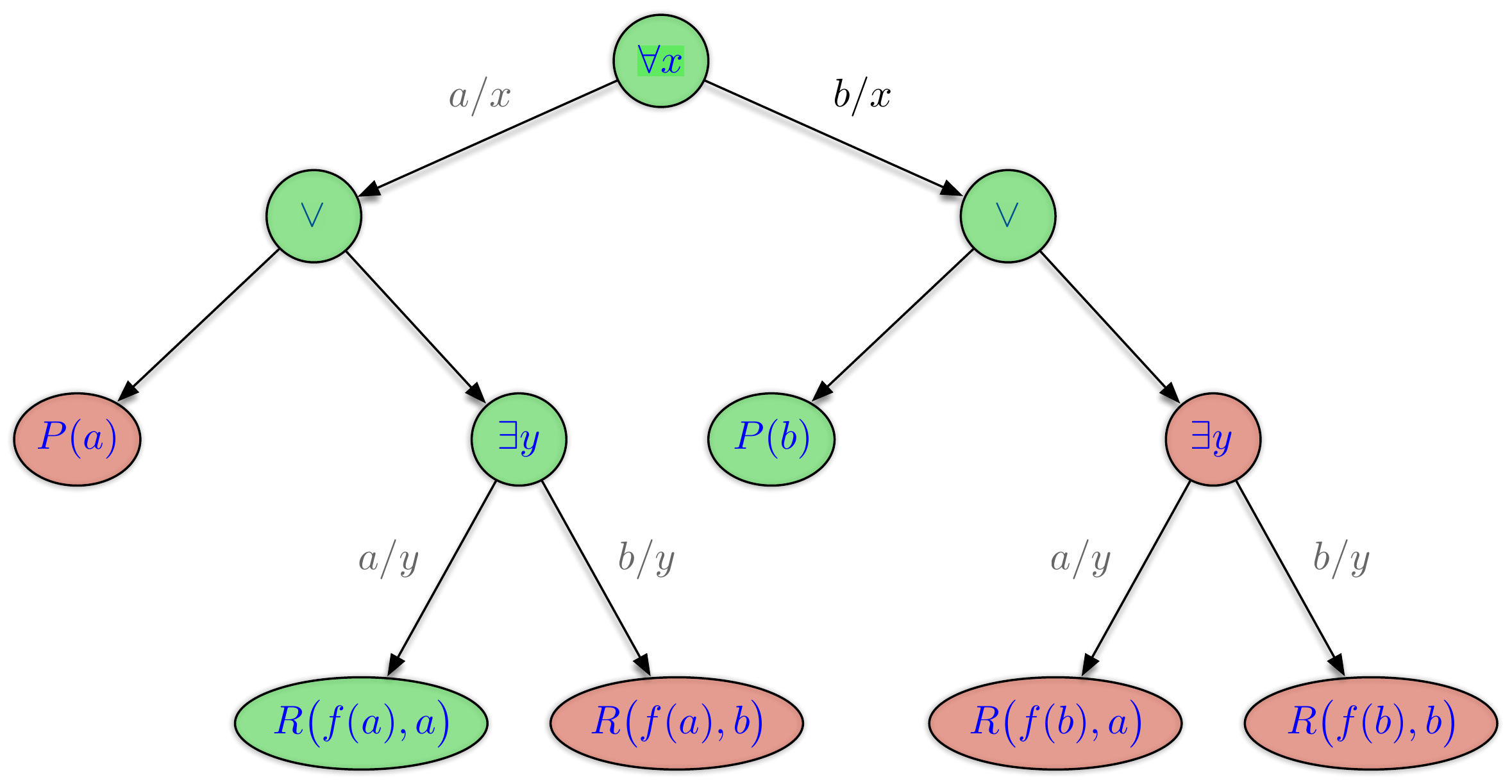}}
\hfill\  \end{figure}
We end up this way with the root being coloured green which shows that the {\em \textbf{V}erifier} has a winning strategy.
 We indicate below by blue arrows such a winning strategy for the {\em \textbf{V}erifier}.
\begin{figure}[H]
\hfill\  
 \shadowbox{\includegraphics[width=8cm]{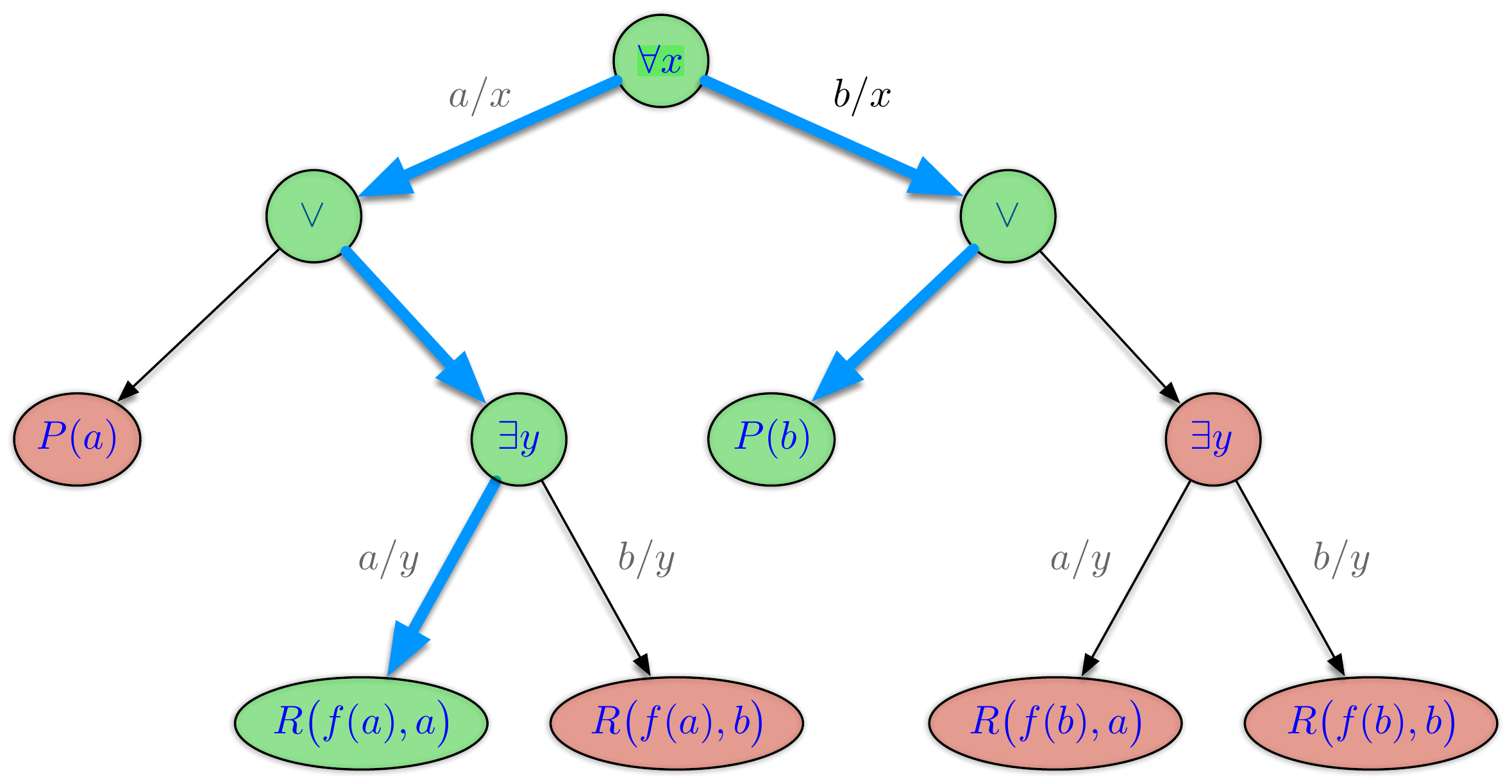}}
\hfill\  \end{figure}
\vspace{-3ex}
\end{example}


\section{The classical proof of \textbf{LS$\downarrow$} with Skolemization}

Given any theory $T$ as in Theorem \ref{DownwardLS}, without loss of generality, one first assumes  that 
every formula ${\phi}\in T$ is in \emph{prenex normal form} i.e. 
$\phi= Q_1 x_1 \ldots Q_k x_k \psi$, where, for all $1\leq i < j\leq k$,  $Q_i ,Q_j \in \{\forall, \exists\}$, $x_i \neq x_j$ and $\psi$ is 
quantifier free. Then, the usual proof of \textbf{(LS$\downarrow$)} goes through the following steps

\begin{enumerate}
\item The  \emph{skolemization} of each such $\phi$, which consists of 
	\begin{enumerate}
		\item first extending the language by adding,  for each existential 
		quantifier that $\phi$ contains, a new function symbol  ${f^{\phi}_k}^{(q_k)}$ of arity $q_k$.
$$\mathcal{L}'= \mathcal{L} \cup \left \{ {f^{\phi}_k}^{(q_k)} \mid Q_k = \exists, q_k= \cardinality{ \big\{ m \mid 1\leq m < k \wedge Q_m= \forall \big\}} \right \}.$$
With this definition $q_k$ coincides with the number of universal quantifiers which precede the existential $Q_k$.
		
		\item   For each existentially quantified variable $x_k$, replacing inside $\psi$, each occurrence of $x_k$  
		by the term\footnote{A function symbol whose arity is zero is simply a constant symbol.} 
			$f^{\phi}_k \left(x_{p_1}, \ldots x_{p_{q_k}}\right)$, where $x_{p_1}, \ldots, x_{p_{q_k}}$ are the universally quantified 
			variables  that precede $Q_k$. 
			More formally:$$t_k=f^{\phi}_k \left(x_{p_1}, \ldots x_{p_{q_k}}\right)$$
$\textup{ where }\{x_{p_1}, \ldots x_{p_{q_k}}\} = \{ m \mid 1\leq m < k \wedge Q_m= \forall \}$ 
			and the sequence of subscripts $(p_i)_{1\leq i\leq q_k}$ is strictly increasing. We then obtain
			\[
			\tilde{\psi}= \psi_{_{\left[{}^{t_{k_1}}{/x_{k_1}}, \cdots, {}^{t_{k_m}}{/x_{k_m}}\right]}}
			\]
			where  $\{x_{k_i}\mid{1\leq i\leq m}\}$ is the set of all existentially quantified variables of $\phi$.
		\item Removing all universal quantifiers from $\phi$. The Skolem normal form of $\phi$ -- denoted $\sigma_{\phi}$ -- becomes:
$$
			\sigma_{\phi}= Q_{i_1} x_{i_1}\ldots Q_{i_t} x_{i_t} \tilde{\psi}
$$
			where the sequence of subscripts $(x_i)_{1\leq i\leq t}$ runs through all universally quantified  variables of $\phi$.
	\end{enumerate}
This way,  any $\mathcal{L}$-theory $T$ is turned into  its skolemized version: some $\mathcal{L}'$-theory $\sigma_{T}=\{\sigma_{\phi}\mid\phi\in T\}$. 

\item One shows that the cardinality of the set  of new function and constant symbols that have been added to the language
 is either countable if the 
language is finite, and it is  the same  as the one of the original language $\mathcal{L}$ if it is infinite. Therefore the extended language $\mathcal{L}'$
 has cardinality $\max \{\cardinality{\mathcal{L}}, \aleph_0\}$.

\item One takes any  $\mathcal{L}$-structure $\mathcal{M}$ of cardinality $\lambda > \kappa$ such that $\mathcal{M}\models T$ and construct some 
$\mathcal{L}'$-structure $\mathcal{M}'$ by extending $\mathcal{M}$ from $\mathcal{L}$ to $\mathcal{L}'$. This is done by providing for every new symbol 
${{f^{\phi}_k}^{(q_k)}}$ an interpretation 
$${{f^{\phi}_k}^{(q_k)}}^{\calmind}: |\mathcal{M}|^{q_k}\mapsto |\mathcal{M}|
~\mbox{ such that it satisfies }~\mathcal{M}' \models \sigma_{T}.$$

With the classical approach, the description of the extension $\mathcal{M}'$ is usually messy, whereas it simply does not exist
 with the game-theoretical approach.

\item \label{iciquatre} One constructs some
 $\mathcal{L}'$-structure  $\mathscr{N}'$ of cardinality $\kappa$ such that $\mathscr{N}'\models\sigma_{T}$ holds. 
So, one selects some sub-domain of   $|\mathcal{M}'|$ which is closed under all interpretations of functions
\footnote{Including the constants which are the ones of arity 0.} of $\mathcal{L}'$.
For this purpose, one starts with any subset $N_0\subseteq|\mathcal{M}'|$ of cardinality $\kappa$ which contains all interpretations of 
constant symbols from $\mathcal{L}'$. We let $\mathscr{F}_k(\mathcal{L}')$ denote the set of  function symbols of  $\mathcal{L}'$ of arity $k$.
	By induction, one defines
$$N_{n+1}=N_n \cup \{ f^{\calmind'}(a_1, \ldots, a_k) \mid k \in \mathbb{N}, f \in \mathscr{F}_k(\mathcal{L}'), a_1, \ldots, a_k \in N_n\},
$$
	and one sets $N_\omega= \bigcup_{n\in \mathbb{N}} N_n$.
	One observes that:
		\begin{enumerate}
			\item\label{iciquatrea} $N_0 \subseteq N_1 \subseteq \ldots \subseteq N_\omega$.
			\item \label{iciquatreb} For all $n \in \mathbb{N},\ \cardinality{N_{n+1}}= \cardinality{N_n}=\kappa$, since the induction yields		
				$$\begin{array}{lll}
				\kappa&\leq &\cardinality{ N_{n}\cup\{ f^{\calmind'}(a_1, \ldots, a_k) \mid k \in \mathbb{N}, f \in \mathscr{F}_k(\mathcal{L}'), a_1, \ldots, a_k \in N_n \} }\\
				\ &\leq &\cardinality{\mathcal{L}' \times N_n^{<\omega}}\\
				\ &\leq &\cardinality{\kappa \times \kappa^{<\omega}}\\
				\ &\leq &\kappa\cdot \kappa\\
				\ &\leq &\kappa
				\end{array}$$
		
			\item\label{iciquatrec} Hence $\cardinality{N_\omega}= \kappa$ holds, since $$\kappa\leq \cardinality{N_\omega}\leq \cardinality{\prod_{n\in\omega}N_n}\leq \cardinality{ \aleph_0\times \kappa} \leq\aleph_0\cdot \kappa 
			\leq \kappa.$$
			\item\label{iciquatred} $N_\omega$ is closed under all interpretations of function symbols from $\mathcal{L}'$. For if  $k\in \mathbb{N}$, $f\in \mathscr{F}_k(\mathcal{L}')$, and $a_1, \ldots, a_k \in N_\omega$,  there would then exist some   $n\in\mathbb{N}$ such that $a_1, \ldots, a_k \in N_n$, henceforth
			  $f^{\calmind'}(a_1, \ldots, a_k) \in N_{n+1} \subseteq N_\omega$.
		\end{enumerate}
		
		Therefore  $\mathscr{N}'$ is defined by:
		\begin{itemize}\renewcommand{\labelitemi}{$\bullet$}
			\item $|\mathscr{N}'|= N_\omega$;
			\item for all constant symbols $c$, $c^{\mathscr{N}'}=c^{\calmind'}$;
			\item for all fonction symbols   $f$ of arity $k$, $f^{\mathscr{N}'}=f^{\calmind'}|_{{N_\omega}^k}$;
			\item for all relation symbols   $R$ of arity $k$, $R^{\mathscr{N}'}=R^{\calmind'} \cap {(N_\omega)}^k$.
		\end{itemize}

\item Finally, one shows that the model $\mathscr{N}$ defined as the restriction of $\mathscr{N}'$ to the language $\mathcal{L}$ satisfies 
 $\mathscr{N}\models T$. This requires once again to review  the construction of $\mathscr{M}'$, by backward this time: 
 another wearisome moment for the students.  

\end{enumerate}


\section{An alternative game-theoretical proof of \textbf{LS$\downarrow$}}

Compared to the previous proof, the game-theoretical proof that we advocate is simpler for it only focuses on the original model and 
fixed winning strategies that witness that the theory holds in this model. Indeed, this proof  contains
\begin{itemize}
  \item no Skolem normal form
  \item no extended language $\mathcal{L}'$
  \item no extension $\mathcal{M}'$ of $\mathcal{M}$ 
  \item no restriction $\mathcal{N}$ of $\mathcal{N}'$ 
\end{itemize}

In this proof, we start with any model  $\mathscr{M}$ such that  both $\cardinality{|\mathscr{M}|}\geq \kappa$ and $\mathscr{M}\models T$ 
hold. We also  assume that every formula $\phi\in T$ is in prenex normal form\footnote{We suppose that $T\neq\emptyset$ holds; 
 otherwise, the result is simply straightforward.}.
\begin{enumerate}
  \item For each $\phi\in T$ we pick any winning strategy $\sigma_{\phi}$ 
  for the {\em \textbf{V}erifier} in $\EV{\mathcal{M}}{{\phi}}$. By looking at the very rules of 
  the game, every student realises immediately that for each existential $Q_ix_i$ in $\phi$, the given strategy secures 
  one element from $|\mathcal{M}|$ that {\em only depends on} the previous choices made by her opponent ({\em \textbf{F}alsifier}). Since 
   $\phi$ is in prenex normal form, these choices made by {\em \textbf{F}alsifier} 
   correspond precisely to the universal quantifiers preceding $Q_ix_i$. For instance, if $\phi$ is of the form 
   $$\forall x_1\exists x_2\forall x_3\forall x_4\forall x_5\psi$$
   then the choices that {\em \textbf{V}erifier}  makes --  following $\sigma_{\phi}$ -- of an element $a_2\in|\mathcal{M}|$ for $x_2$ and an element 
   $a_5\in|\mathcal{M}|$ for $x_5$ depend respectively of the choices made by {\em \textbf{F}alsifier} of an element  $a_1\in|\mathcal{M}|$ for $x_1$, 
   and of elements  $a_1\in|\mathcal{M}|$ for $x_1$, $a_3\in|\mathcal{M}|$ for $x_3$, $a_4\in|\mathcal{M}|$ for $x_4$. 
   In other words, the winning strategy picks for $x_i$ an element that is {\em function of } -- meaning that it only depends on -- the choices made for the universally quantified 
   variables that come before $x_i$. Assuming there is $k$-many such  universally quantified 
   variables, this induces a unique function $f_i^{\sigma_\phi}: |\mathscr{M}|^{k}\mapsto |\mathscr{M}|$.
     So we come up with a set  $\mathscr{F}=\left\{f_i^{\sigma_\phi}\mid \phi\in T\right\}\cup\left\{f^{\calmind}\mid f\in\mathcal{L}\right\}$ of functions of different 
   arities\footnote{Functions of arity 0  being identified with elements of the domain 
   $|\mathscr{M}|$.}  whose cardinality is at most $\kappa= \max \{\cardinality{\mathcal{L}}, \aleph_0\}$.
    
  \item We take any subset $N_0\subseteq |\mathscr{M}|$ of cardinality $\kappa$ and  proceed as in (4)(a-d)
  to obtain  the least (for inclusion) subset 
  $N\subseteq |\mathscr{M}|$ of cardinality $\kappa$
   that satisfies both  $N_0\subseteq N$ and $N$ is closed under all functions in $\mathscr{F}$.  
  
  \item We form $\mathcal{N}$ as the restriction of $\mathcal{M}$ from $|\mathcal{M}|$ to $N$, and show that $\mathcal{N}\models T$ in a straightforward manner this 
  time, since   for every formula $\phi\in T$ the very same strategy\footnote{Strictly speaking this is the 
  restriction of this strategy to $N$.} $\sigma_{\phi}$ which is winning for  the {\em \textbf{V}erifier} in $\EV{\mathcal{M}}{{\phi}}$ is 
  also winning for the {\em \textbf{V}erifier} in $\EV{\mathcal{N}}{{\phi}}$. 
\end{enumerate}

\section{An even simpler proof of \textbf{LS$\downarrow$}}

This proof does not even  require to go through the formulas of $T$ to be in prenex normal form. 

\begin{enumerate}
  \item for each  $\phi\in T$, pick any winning strategy $\sigma_{\phi}$  for the {\em \textbf{V}erifier} in $\EV{\mathcal{M}}{{\phi}}$ and for 
  any $A\subseteq|\mathcal{M}|$ consider all possible plays in $\EV{\mathcal{M}}{{\phi}}$ such that \begin{enumerate}
  \item {\em \textbf{F}alsifier} restricts his $\forall$-moves to choosing  elements of $A$, and 
  \item {\em \textbf{V}erifier} applies her winning strategy $\sigma_{\phi}$.
\end{enumerate} 
 set  $(A)^{\sigma_{\phi}}$ as the subset of  $|\mathcal{M}|$ formed of all the $\exists$-moves made by {\em \textbf{V}erifier}.
  Set also  
  $(A)^{\mathscr{F}}=\left\{f^{\calmind}(\vec{a})\mid f\in\mathcal{L} \textit{ a function symbol with arity }k, \vec{a}\in {N^{\mathbf{V}}_{n}}^{k}\right\}$,\item   inductively 
define $N$ by:
\begin{enumerate}
  \item $N^{\mathbf{F}}_0\subseteq|\mathcal{M}|$ any set s.t. $\cardinality{N^{\mathbf{F}}_0}=\kappa$, $\{c^{\calmind}\mid c \textit{ constant }\in\mathcal{L} \}\subseteq N^{\mathbf{F}}_0$, 
  \item $N^{\mathbf{F}}_{n+1}=N^{\mathbf{F}}_n\cup\left(N^{\mathbf{F}}_n\right)^{\mathscr{F}}\cup{\displaystyle\bigcup_{\phi\in T}}
\left(N^{\mathbf{F}}_n\right)^{\sigma_{\phi}}$, and 
 $N={\displaystyle\bigcup_{n\in\mathbb{N}}}N^{\mathbf{F}}_n$.  Then,
\begin{multicols}{3}
\begin{itemize}
  \item $\cardinality{N}=\kappa$
   \item $(N)^{\sigma_{\phi}}\subseteq N$
   \item $(N)^{\mathscr{F}}\subseteq N$.
\end{itemize}
\end{multicols}

\end{enumerate}
  \item Form $\mathcal{N}$ as the restriction of $\mathcal{M}$ to $N$, and easily verify that $\mathcal{N}\models T$, since the very same 
$\sigma_{\phi}$  is winning for  {\em \textbf{V}erifier} in $\EV{\mathcal{N}}{{\phi}}$.
\end{enumerate}

\section{Conclusion}We tried these proofs on our own students at EPFL. It turns out that they 
understand much better the proof of  {LS$\downarrow$} that we recommend than they buy the classic one. It also requires 
much less time to teach, and above all we deeply believe that this proof highlights 
what is essential in this result now devoid of all the technical details of the skolemization.  On the other hand, there is a price to pay in 
doing so: one has to present the semantics of first-order logic through evaluation games. But here also we have noticed that 
the students learn more easily this other part of the course. Another advantage of the game-theoretical approach 
is also that it paves the way for 
the {\em back-and-forth} method \cite{fraisse1950nouvelle} 
or the Ehrenfeucht-Fra\"{i}ss\'{e} games \cite{ehrenfeucht1961application} that are intensively used in model theory
\cite{marker2002model}.

\bibliography{tools_for_teaching_logic_2015}

\end{document}